\documentclass[twocolumn,aps,pre,showkeywords,preprintnumbers]{revtex4}
\pdfoutput=1
\usepackage[dvipdf]{graphicx}
\usepackage{epstopdf,epsfig,subfigure}
\usepackage{hyperref,graphics,color}
\usepackage{amsmath,mathrsfs,dcolumn}
\usepackage{latexsym,amssymb}
\begin{document}
\title{ Stochastic resonance and first passage time for excitable system exposed to underdamped medium}
\author{Solomon Fekade Duki}
\affiliation {National Center for Biotechnology Information,
National Library of Medicine and National Institute of Health,
8600 Rockville Pike, Bethesda MD, 20894 USA}
\author{Mesfin Asfaw Taye}
\affiliation{West Los Angeles College, Science Division, Department of Physics
9000 Overland Ave, Culver City, CA 90230}

\begin{abstract} 
Noise induced Brownian dynamics in underdamped medium is studied numerically to understand the 
firing time of excitable systems. By considering  Brownian particles that move in underdamped medium,
we study how the first arrival time behaves for different parameters of the model. We study 
the first arrival time for both single particle as well as the first arrival time of one particle 
out a system that has $N$ particles. The present study helps to understand the 
intercellular calcium dynamics in cardiac tissue at the level of a single microdomain and at
a tissue level (ensemble of microdomains). In the presence of time varying signal, we study
how signal to noise ratio (SNR) depends on the model parameters. It is showed that the SNR 
exhibits a pronounced peak at a particular noise strength.  The fact that the SNR is amplified 
as the number of micro domains ($N$) increase shows that the weak periodic signal plays a decisive 
role in controlling the noise induced dynamics of excitable systems which may also shed light on 
how to control the abnormal calcium release in a cardiac tissue.

\end{abstract}
\pacs{Valid PACS appear here}
\maketitle

\section {Introduction}

Understanding the physics of thermally activated barrier crossing is vital since  it has diverse 
physical applications and serves as a tool to understand  stochastic paradigms. Particularly if 
one considers a Brownian particle moving in a viscous medium, assisted by the thermal background 
kicks, the particle crosses the potential barrier. The magnitude of its first passage time relies
not only on the system parameters, such as the potential barrier height, but also on the initial 
and boundary conditions.  Understanding of such noise induced thermally activated 
barrier crossing problem  is vital  to get a better understanding of most biological problems such
as cardiac system \cite{am1,am2,am3,am4,am5,am6,am7,am8}. In the past, considering temperature 
independent viscous friction, the dependence of the mean first passage time (equivalently
the escape rate) on model parameters has been explored for  various model systems \cite {am9}. 
However experiment shows that the viscous friction $\gamma$ is indeed temperature dependent and 
it  decreases as temperature increases. In this work we discuss the role of temperature on the 
viscous  friction as well as on the first passage time  by taking a viscous friction $\gamma$ that 
decreases exponentially when the temperature $T$ of the medium increases  ($\gamma=Be^{-A T}$) as 
proposed originally by Reynolds \cite{am10}.

Exposing  excitable systems to time varying periodic  forces may result in an intriguing 
dynamics where in this case the coordination of the noise with time varying force leads to 
the phenomenon of stochastic resonance (SR) \cite{am13,am14}, provided that the noise induced 
hopping events synchronize with the signal. The phenomenon of stochastic resonance (SR) 
has obtained considerable interests because of its significant practical applications in 
a wide range of fields. SR depicts that systems  enhance their performance as long as   
the thermal background noise  is synchronized with time varying  periodic signal. Since 
the innovative work of Benzi {\em et. al.} \cite{am13}, the idea of stochastic resonance has been 
broadened and implemented in many model systems \cite{am15,am16,am17,am18,am19,am20,am21,am22,am23}.  
Recently the occurrence  of stochastic resonance for a Brownian particle as well as 
for extended system such as polymer has been reported by us \cite{am24,am25}.  Our analysis 
revealed that, due to the flexibility that can enhance crossing rate and change in chain 
conformations at the barrier, the power amplification exhibits an optimal value for optimal 
chain lengths  and elastic constants  as well as for optimal noise strengths. However 
most of these studies considered a viscous friction which is temperature independent. 
In this work considering temperature dependent viscous friction, we study how the 
signal to noise ratio (SNR) behaves as one varies the model parameters. We explore first the 
stochastic resonance (SR) of a single particle (equivalently of a single microdomain) and 
later we study the SR for many particle system ($N$ microdomains) by considering both 
temperature dependent and  independent viscous friction and compare the result. 

The aim of this paper is to explore the crossing rate and stochastic resonance of a 
single as well as many Brownian particles in an underdamped medium. 
Although a generic model system is considered,  the present study  also helps to 
understand the dynamics of abnormal calcium cycle in a cardiac system since the 
thermally activated barrier crossing rate of a single particle mimics the dynamics  
of abnormal calcium flow at a single microdomain level while the dynamics for many 
particles case is related to the flow of calcium at tissue (many microdomains) level 
\cite {am11,am12}.

To give a brief outline, in this work we first study the first passage time of a 
single particle for both temperature dependent and independent viscous friction cases. 
The  numerical simulation results depict that the first passage time is smaller when $\gamma$ 
is temperature dependent.  In both cases the escape rate increases as 
the noise strength increases and  decreases as the potential barrier increases.  We 
then  extend our study for $N$ particle systems. The first passage time $T_{N}$ that 
one particle (out of $N$ particles) takes to cross the potential barrier can be studied 
via numerical simulation. It is found that $T_{N}$ is considerably  smaller when the 
viscous friction is temperature dependent. For both cases, $T_{N}$ decreases as the noise
strength increases and as the potential barrier steps down.  In high barrier limit, 
$T_{N}=T_{s}/N$ where $T_{s}$ is the first arrival time  for a single particle. In 
general as the number of particles $N$ increases, $T_{N}$ decreases.

We then study our  model system in the presence of time varying signal. In this case 
the interplay between noise and sinusoidal driving force in the bistable system may 
lead the system into stochastic resonance. Via numerical simulations, 
we study how the signal to noise ratio (SNR) behaves 
as a function of the model parameters. The SNR depicts a pronounced peak
at particular noise strength $T$. The SNR is higher for temperature
dependent $\gamma$ case. In the presence of $N$ particles,  SNR is considerably 
amplified as $N$ steps up showing the  weak periodic signal plays a vital   
role in controlling the noise induced dynamics of excitable systems.

The rest of the paper is organized as follows. In section II, we present the model. 
In section III, by considering both temperature dependent and independent viscous 
friction cases, we explore the dependence for first arrival time on model parameters for both 
single as well as many particle system. The role of  sinusoidal driving force on 
enhancing the mobility of the particle is studied in IV. Section V deals with 
Summary and Conclusion.

\section{The model}
We consider a Brownian particle that moves in underdamped medium. The particle is exposed
to a piecewise linear potential with a reflective boundary on the left edge and an absorbing
boundary on the right. The potential is given by 
\begin{equation}
  U(x)=\left\{
  \begin{array}{ll}
	    -U_{0}\left({~2x\over L_{0}}+1\right),& \text{if}~~~ x < -L_{0};\\ 
    U_{0}\left({~2x\over L_{0}}+1\right),& \text{if}~~~ -L_0 \le x \le {0};\\ 
    U_{0}\left(-{2x\over L_{0}}+1\right),& \text{if}~~~{0} \le x \le L_{0}\\ 
		    U_{0}\left({~2x\over L_{0}}-1\right),& \text{if}~~~ x > L_{0}. \\ 
   \end{array}
   \right.
\end{equation}
where  $U_{0}$ and $2L_{0}$ denote the height and width of the barrier, respectively. The 
potential exhibits its maximum value $U_{0}$ at  $x={0}$ and its minima at $x=-L_0$  and 
$x=L_{0}$. The ratchet potential is coupled with a uniform temperature $T$ as shown in Fig. 1.  
\begin{figure} 
\includegraphics[width=7cm]{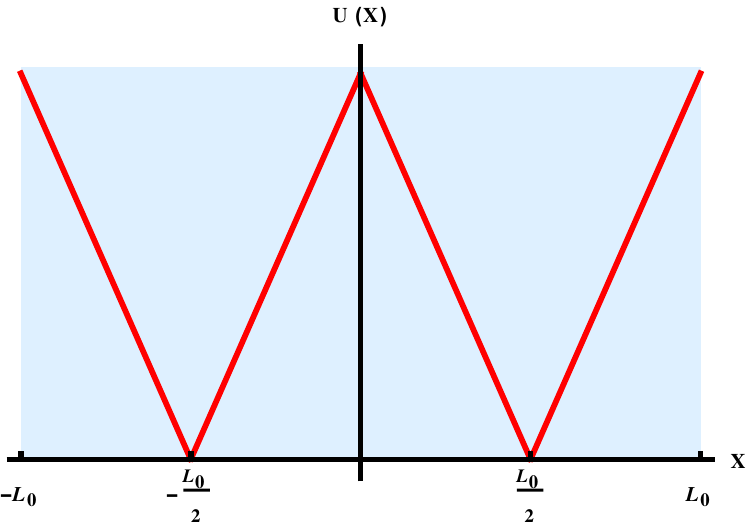}
\caption{ Schematic diagram for a Brownian particle in a piece-wise linear  potential. 
}
\end{figure}

For a Brownian particle that is arranged to undergo a random walk in an underdamped medium, 
the dynamics of the particle is governed by the Langevin equation
\begin{equation}
m{\frac{d^2x}{dt^2}} = - {d U(x) \over dx} -\gamma{dx\over dt} + \sqrt{2k_{B}\gamma T}\xi(t)
\end{equation}
where $m$ is the mass of the particle and $k_B$ is the  Boltzmann's constant. The viscous 
friction $\gamma$ is assume to have an exponential temperature dependence as 
 \begin{eqnarray}
 \gamma & =Be^{-A T},& \text{if}~~~ -L_0 \le x \le {L_{0}}.
\end{eqnarray}
Here $A$ and $B$ are constants that characterize the system, and for brevity we will work
on units where $m$ and $k_B$ are unity.  The background thermal noise $\xi(t)$ 
is assumed to be Gaussian and has no correlation; {\em i. e.} 
$\left\langle \xi(t) \right\rangle =0$ and $\left\langle \xi(t)  \xi(t')\right\rangle=\delta(t-t')$.

As discussed in the work \cite{am11,am12}, the timing of spontaneous excitations in a cardiac  tissue  can be studied via  master equation. For long one dimensional cable (where $N$ is  large enough), the voltage fluctuations can be approximated by Gaussian statistics. In other words, if the number of ensembles $N$ is  large enough,  the calcium dynamics can be well approximated by the corresponding Fokker-Planck type of equation  where the effective potential is very complicated and model dependent as discussed in our previous work \cite{am11}.  Hence the piece-wise linear potential presented in this work only qualitatively addresses the cardiac problem.  Moreover, since  different excitable systems  have different effective potential, the simplified potential presented in this work helps to understand these model systems at least qualitatively. 

Please note that the dynamics of such a system can be realized experimentally. One
makes negatively charged particle, then put the particle within
positively and negatively charged fluidic channel to mimic the piecewise linear potential. The
fluidic channel is subjected to an external periodic force (AC field). Since the
particle is negatively charged, it encounters a difficulty of crossing through
the negatively charged part of the channel. Assisted by the thermal background
kicks along with its conformational change, the particle ultimately
overcomes the barrier. The presence of time varying force, may further enhance
the rate of crossing.

To simplify the numerical simulation we rescale position, time, temperature and barrier potential and work
in terms of dimensionless quantities; length ${\bar x}=x/L_{0}$ , time ${\bar t}=t/\beta$, 
barrier height ${\bar U_{0}}=U_{0}/T$, and temperature $\beta=B L_0^2/T$. 
For brevity we drop all the bars from all equations from here on.

\section{The first arrival time}

In this section, the dependency of the first arrival time on the different model parameters is 
explored via numerical simulations by integrating the Langevin equation (2) (employing Brownian
dynamics simulation).  In the simulation, a Brownian particle is initially situated in one of 
the potential wells. Then the trajectories for the particle   is simulated by considering different
time steps $\Delta t$ and time length $t_{max}$. In order to ensure the numerical accuracy $10^{9}$
ensemble averages have been obtained.

Before exploring the first arrival time of one particle out of $N$ particles, we first numerically
evaluate the first passage time distribution both for a single particle system and many particle systems. 
This gives us a qualitative clue on  how the first passage time  behaves because the first passage
time is given by $T=\int_{0}^{\infty} t' P(t')dt'$  where $P(t')$ is the first time distribution. 

The first passage time distribution $P_{N}(t)$ as a function of $t$ is shown in Fig. 2 for fixed 
values of $U_{0}=2.0$ and  $L_{0}=1.0$. Figure 2(a) shows the distributions for temperature independent
$\gamma$ case ($A=0$) and while Fig. 2(b) shows temperature dependent $\gamma$ case ($A=1.0$).
The simulation results show that when the number of particles $N$ in the system  increases, the peak of the distribution 
decreases, revealing that the firing time for one particle (out 
of the $N$ particles) decreases when $N$ increases.
\begin{figure}
\centering
\subfigure [] 
{
\includegraphics[width=5cm]{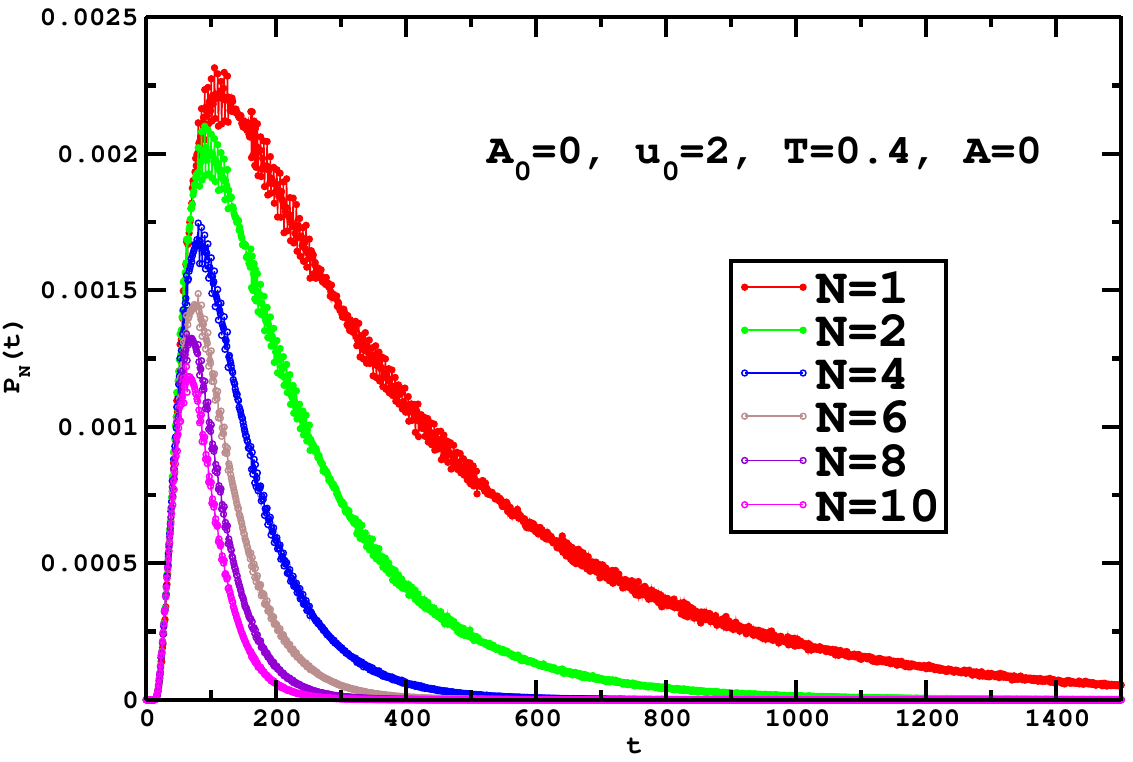}
\vspace{-2\baselineskip}
}
\subfigure [] 
{
\includegraphics[width=5cm]{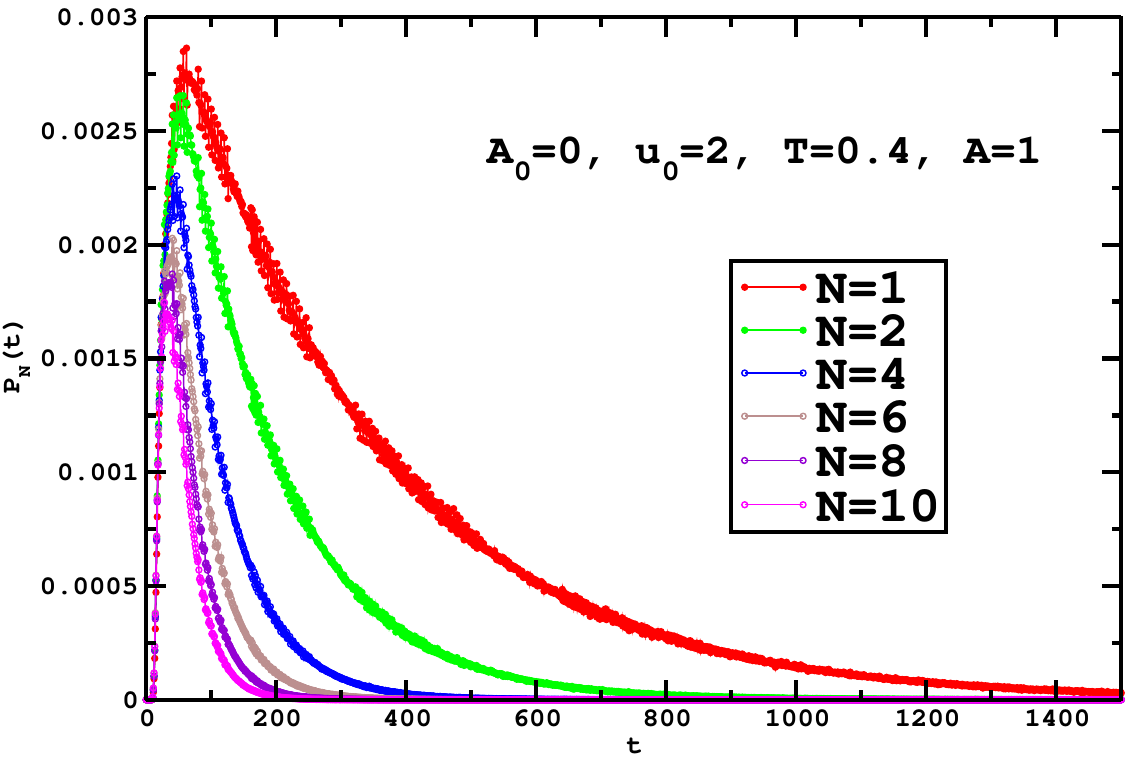}
}
\caption{ (Color online) The first passage time  distribution  $P_{N}(t)$  as a 
function of $t$ for different values of  $N$. We use parameter values of $U_{0}=2.0$ and 
$L_{0}=1.0$. Figures (a) and (b) show the distributions for constant $\gamma$ ($A=0$) and 
temperature dependent $\gamma$ ($A=1$) cases, respectively.}
\label{fig:sub} 
\end{figure}

Note that  in the high barrier limit, the first passage time distribution $P(t)$ is given by 
\begin{equation} 
P_s(t)={e^{{-t\over T_{s}}}\over T_s}
\end{equation}
where $T_s$ is the first passage time for a single particle. For a system that has $N$ particles, the
distribution of first passage time for one of the particles to cross the barrier is given as 
\begin{equation} 
P_N(t)={e^{{-t\over T_{N}}}\over T_N}.
\end{equation}
The first arrival time $T_N$, {\em i. e. } the time for one of the particles first to cross the 
potential barrier, is given by 
\begin{equation}
T_N={T_s\over N}.
\end{equation}

\begin{figure}[]
\centering
\subfigure[] 
{
\includegraphics[width=5cm]{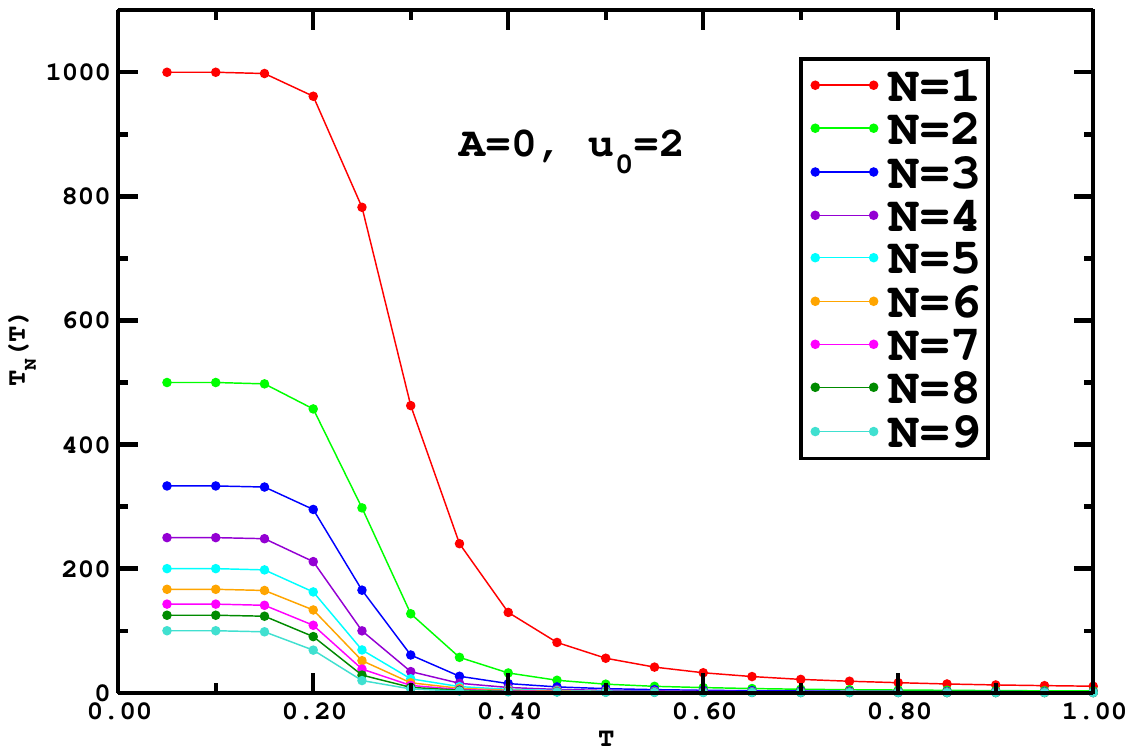}
}
\subfigure[] 
{
    \includegraphics[width=5cm]{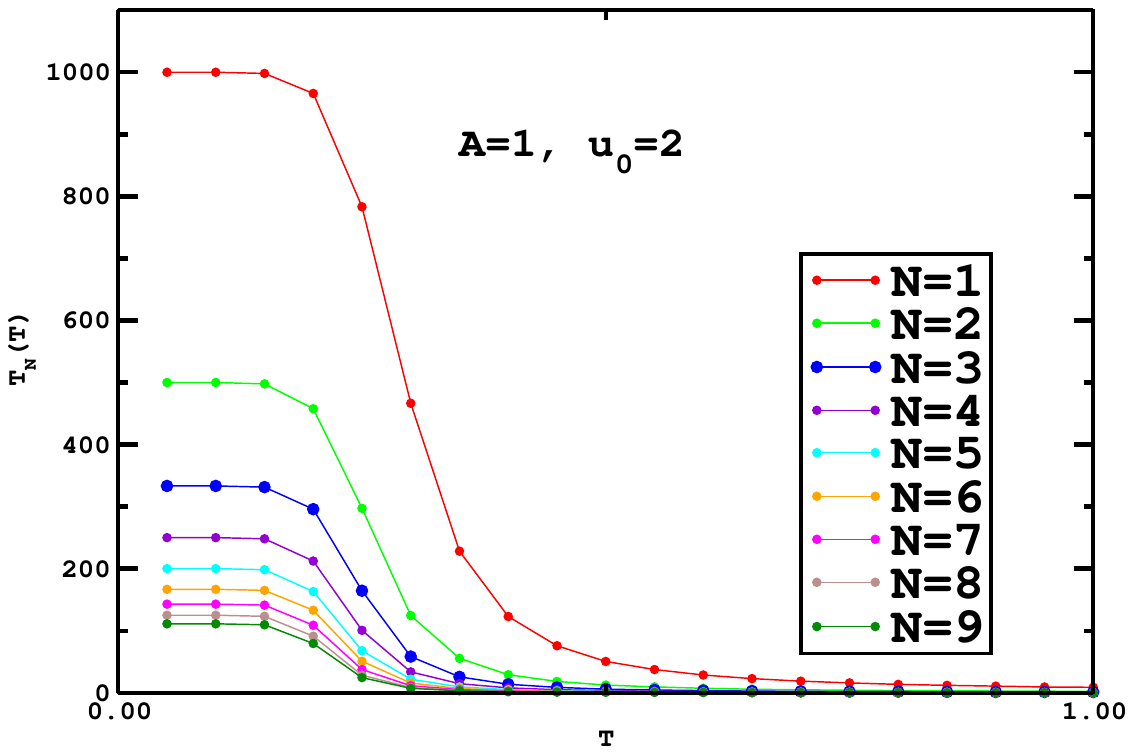}
}
\caption{ Color online) The mean first passage time $T_N$  as a function of temperature $T$ for the 
parameter values of $U_0=2.0$ and $L_{0}=1.0$ for different $N$ values. Figs. 3(a) and 3(b)
are plotted by considering constant and variable $\gamma$ cases, respectively. 
} 
\label{fig:sub} 
\end{figure}

As we discussed before, exploring  noise induced thermally activated barrier crossing is vital
to get a better understanding on how the escape rate or equivalently the first arrival time depends 
on the different model parameters. If one considers a Brownian particle moving in underdamped medium
assisted by the background thermal  kicks, the particle crosses the potential barrier after some time.
The magnitude of its first arrival time relies not only on the system parameters, such as the potential
barrier height, but also on the initial and boundary conditions.

Most previous studies considered temperature independent viscous friction. However, experiment shows 
that viscous friction $\gamma$ is indeed temperature dependent and
it decreases as temperature increases\ref{am26,am27}. Using constant and exponential 
dependence of $\gamma$ on temperature, we have plotted the first passage time $T_N(T)$ 
as a function of temperature  $T$ in Figs. (3a) and (3b). The first passage time decreases as $N$ and 
the strength of the temperature step up in both cases. In the figures the parameters are fixed 
as $U_0=2.0$ and $L_{0}=1.0$. The 
the mean first passage time is lower in the variable $\gamma$ case.
This can be more appreciated if one plots the ratio of the first passage time between cases where $A=1$
and $A=0$ ($\bar{T}_N$) as shown in Fig. 4. In the figure, the ratio is flat at low temperature and increases very
rapidly at higher temperature $T$. 
This is plausible since the diffusion constant 
$D=T/\gamma= k_{B}T e^T$  is also valid when viscous friction to be temperature dependent
showing that the effect of temperature on the particles' mobility is twofold. First, it 
directly assists the particles to surmount the potential barrier; {\em i. e.} particles 
jump the potential barrier at the expenses of the thermal kicks. Secondly, 
when temperature increases viscous friction gets attenuated and hence diffusibility of 
the particle increases. Since  most biological systems operate at physiological temperature, 
our study suggests that macromolecules in such systems are transported fast when molecules 
exposed at temperature dependent $\gamma$.

\begin{figure} 
\includegraphics[width=8cm]{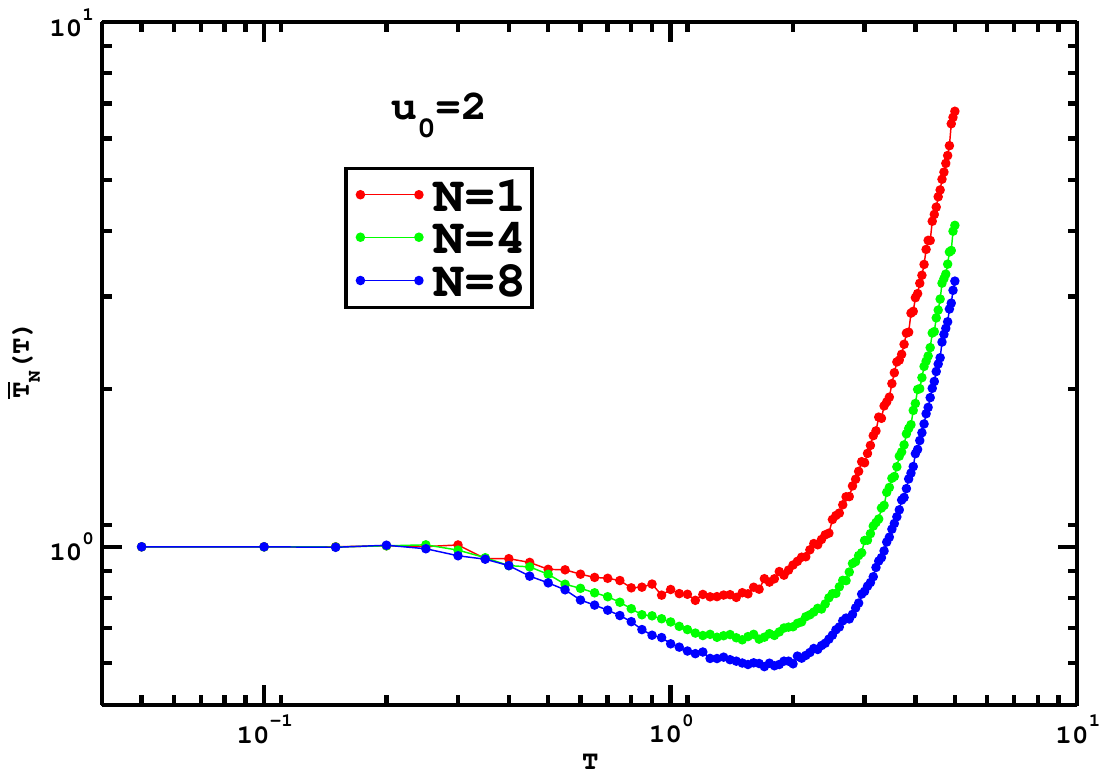}
\caption{ The ratio of first passage time between the temperature dependent $\gamma$ ($A=1$) and constant
$\gamma$ ($A=0$) cases. 
}
\end{figure}

 Furthermore,  when $N$ increases the first arrival time decreases. This suggests that 
in cardiac tissue, when the number of microdomains increases the chance for cardiac 
tissue to release calcium abnormally increases. This is because spontaneous calcium
release at microdomain level triggers  diffusion of calcium in  the  neighboring domains 
and hence causes calcium release at tissue level. Hence increasing the number of microdomains
increases the chance for one of the domain to fire calcium steps up, resulting in reduced 
mean arrival time.

\section{Stochastic resonance }  

In the past,  various studies have shown that exposing excitable systems to time varying periodic 
signals may result in coordination between the noise and the signal to leads the phenomenon of
stochastic resonance (SR) \cite{am13,am14}. This coordination brings SR provided that the noise
induced hopping events synchronize with the time varying periodic signal. To quantify this resonance in
the presence of varying $\gamma$, in this section we study how the signal noise ratio (SNR) behaves 
as function of different model parameters.

In the presence of a time varying periodic signal $A_0 \cos(\Omega t)$, the Langevin equation 
that governs the dynamics of the system is written as  
\begin{equation}
m{\frac{d^2x}{dt^2}} = - {d U(x) \over dx} -\gamma{dx\over dt} + A_0 \cos(\Omega t)+ \sqrt{2k_{B}\gamma T}\xi(t).
\end{equation}
where $A_0$ and $\Omega $ are the amplitude and angular frequency of the external signal, respectively. 
We have numerically simulated eq. 7 for small barrier height and explore the dependence of the first 
passage time distribution $P_N(t)$ in the presence of time varying signal ($A_0 \cos(\Omega t)$).

The first passage time distribution function $P_N(t)$ as a function of time is plotted in Fig. 5
for fixed values of $U_{0}=2.0$ and $T=0.4$ considering  a constant $\gamma$ ($A=0$). Figs. 5a, 5b and 5c show 
the change in the distribution  for $N=1$, $N=6$ and $N=10$, respectively. As a comparison,  we plot the first time distributions both in the presence of the external
signal $A_0=1.0$ (green solid line) and in the absence of signal $A_0=0.0$ (red solid line). As it can be seen 
clearly, the presence of external time varying signal cases multiple resonances.
. 
\begin{figure}[]
\centering
\subfigure [ ] 
{
  \includegraphics[width=5cm]{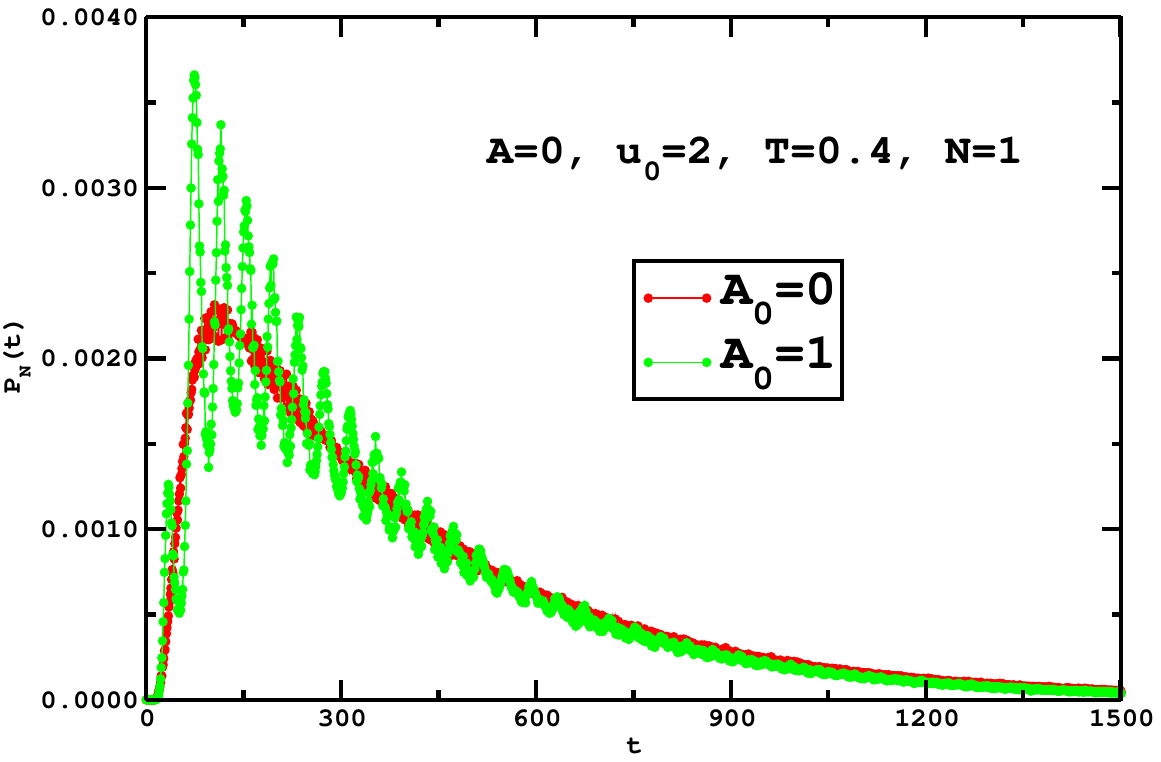}
}
\subfigure[ ] 
{
  \includegraphics[width=5cm]{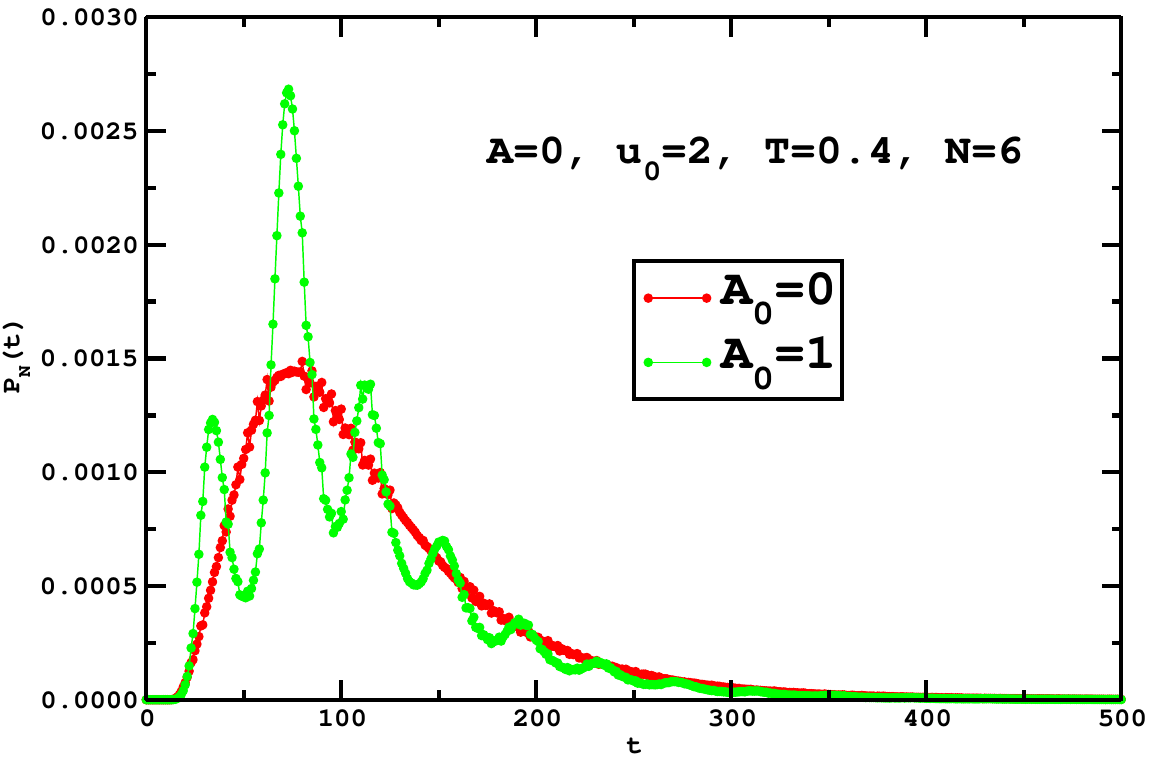}
}
\subfigure [ ] 
{
  \includegraphics[width=5cm]{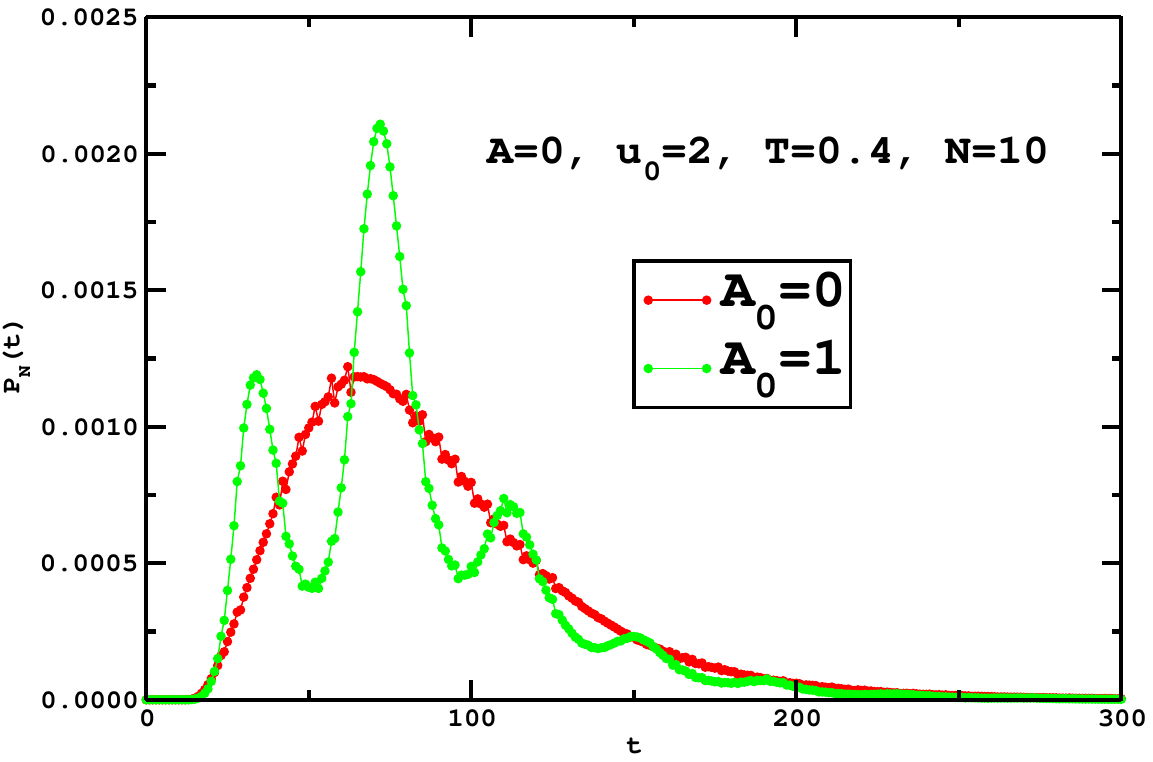}
}

\caption{ (Color online) The first passage time  distribution  $P_N(t)$  as a function of time 
for the parameter values of $U_0=2.0$, $\Omega=0.4$, and $A_0=0.1$.  The red and green lines 
show the case where external signal is turned on and off, respectively. In Figs. 5(a), 5(b) 
and 5(c),  $N$ is fixed at $N=1$, $N=6$ and $N=10$, respectively.} 
\label{fig:SR1}
\end{figure}

The resonance profile can be better observed by looking at the ratio between the first passage time 
distribution functions in Fig. 6, namely by plotting the ratio of the green curve to the red one.
Fig. 6a shows the case for $N=1$ and it clearly shows the resonance profile. It turns out that the
ratio of the first time distribution is independent of the number of particles as shown in Fig. 6b
where we plot  the ratio for different $N$ values varying from 1 to 10. \\

\begin{figure}[]
\centering
\subfigure[ ] 
{
    \label{fig:sub:a}
    \includegraphics[width=5cm]{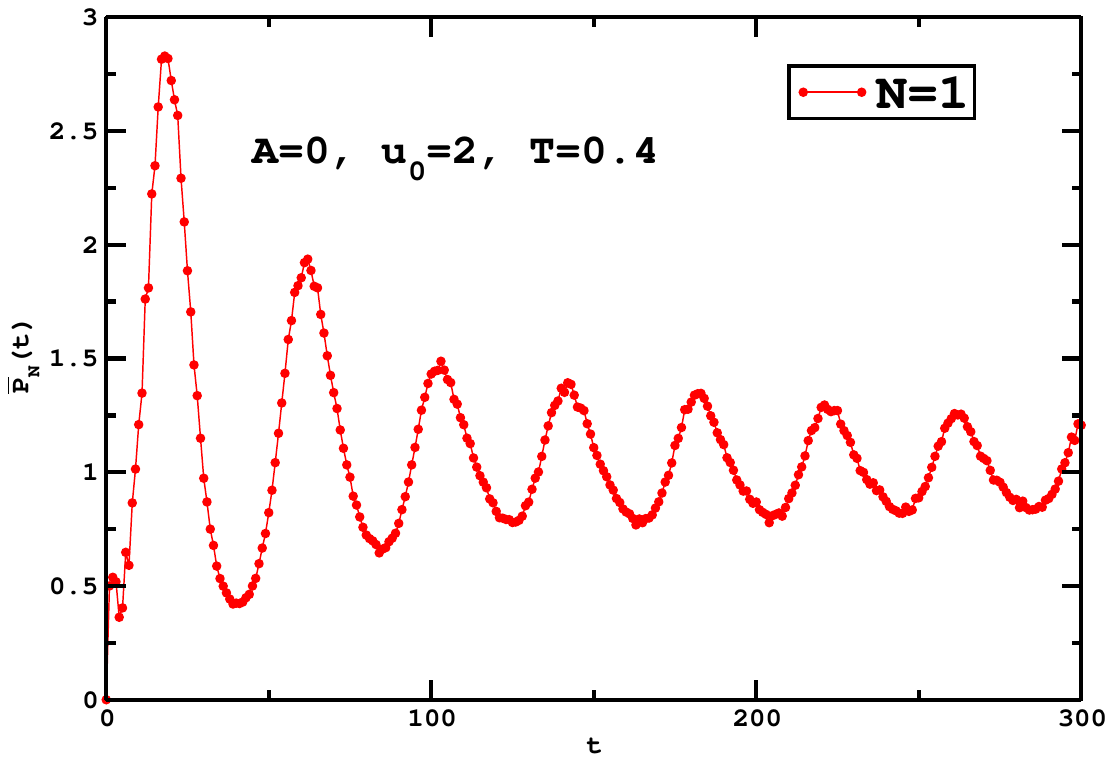}
}
\subfigure[ ] 
{
    \label{fig:sub:b}
    \includegraphics[width=5cm]{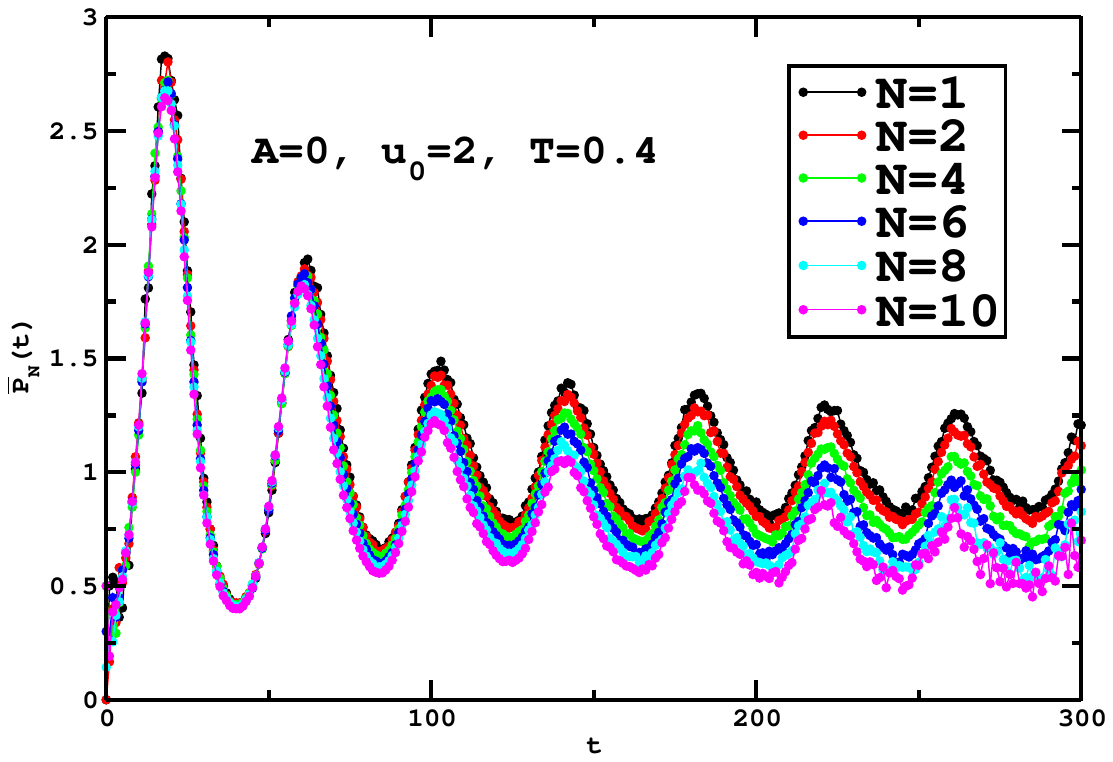}
}
\caption{ (Color online) ) The ratio of the first passage time distribution functions 
$\bar{P}_{N}(t)$ when the signal amplitude is $A_0=0.1$ and frequency is $\Omega=0.4$. In Fig. 6(a), the particle number is fixed as $N=1$. In Fig. 6(b), the resonance profile is independent on $N$.
} 
\label{fig:SR_ratio}
\end{figure}

In the limit of small 
barrier height, we also plot  the first passage time
distribution in Fig. 7. In Figs. 7a, 7b and 7c, the number of particles is fixed as $N=1$, $N=6$ and $N=10$, respectively. 
The figures show that the resonance is more pronounced when $\gamma$ is
temperature dependent (green line) than temperature independent $\gamma$ (red line).

\begin{figure}[]
\centering
\subfigure[] 
{
  \includegraphics[width=5cm]{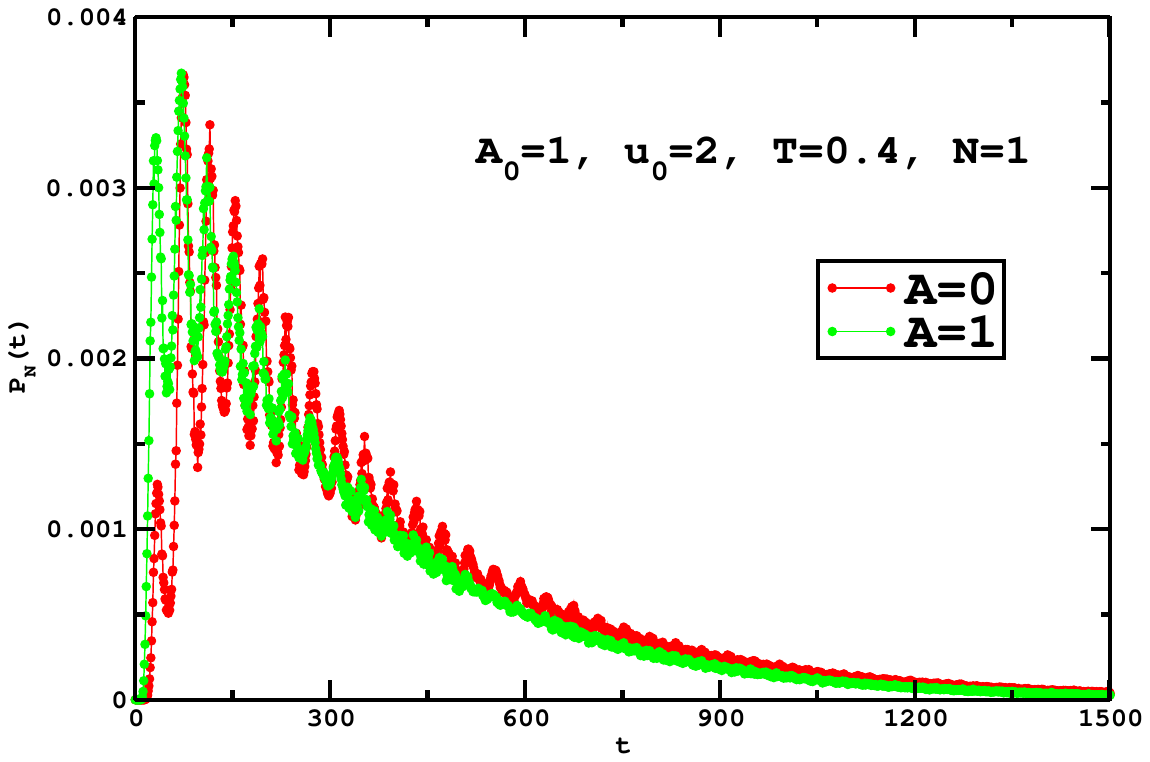}
}
\subfigure[] 
{
  \includegraphics[width=5cm]{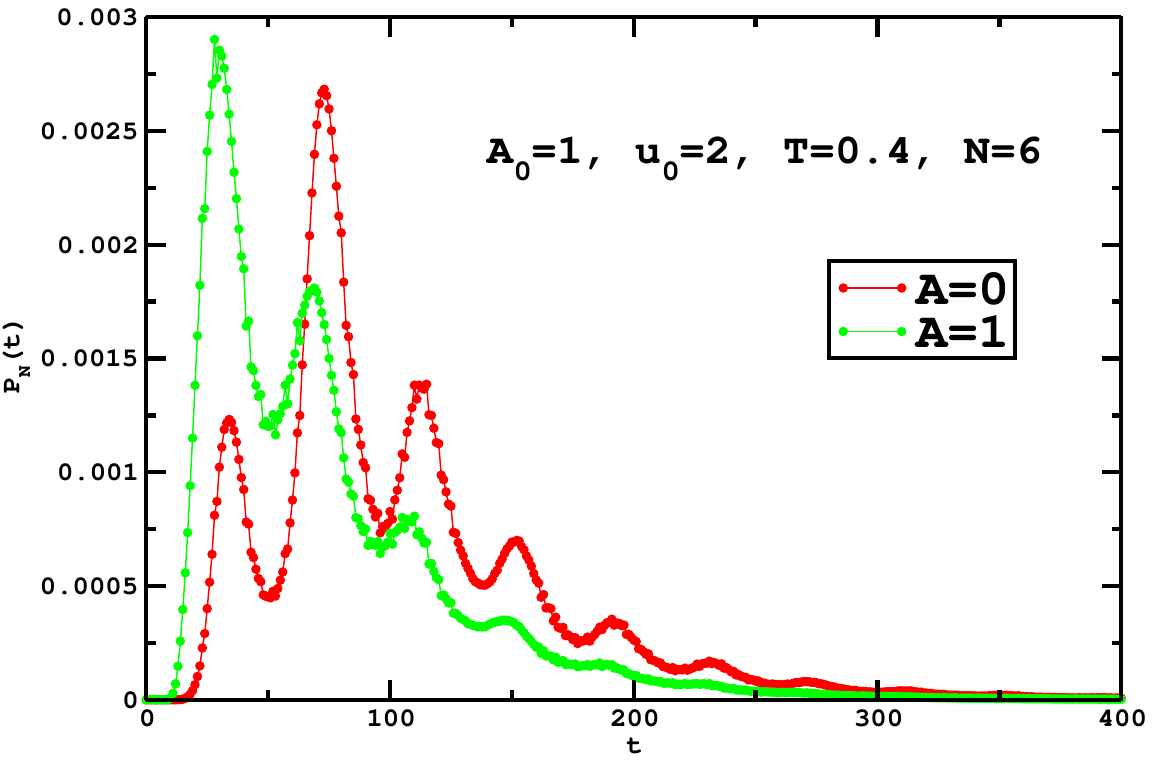}
}
\subfigure[] 
{
  \includegraphics[width=5cm]{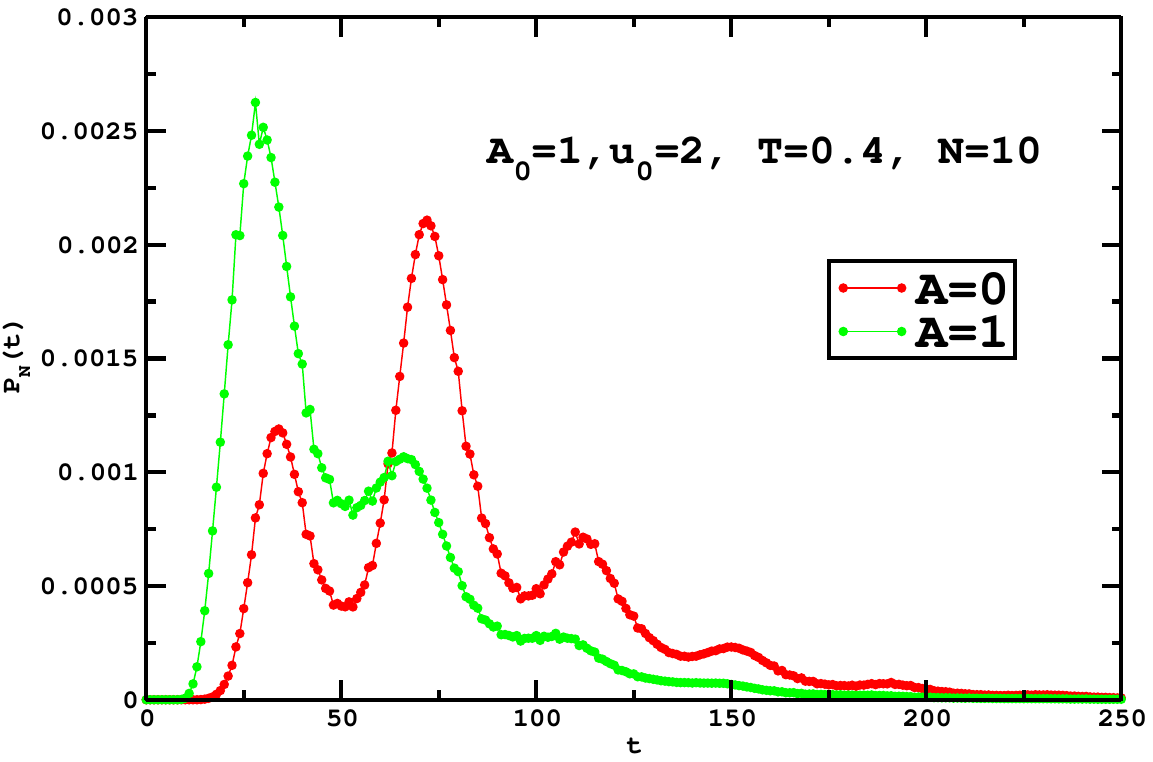}
}

\caption{ (Color online) The  first passage time distribution $P_N(t)$  as a function of time 
in small  barrier limit.  Figs. (a), (b) and (c) are plotted  by considering one, six and ten
particles cases respectively. The amplitude and frequency of the signal are set at $A_0=0.1$
and $\Omega=0.4$ respectively.
} 
\label{fig:sub} 
\end{figure}

Via numerical simulations, we further study how the SNR behaves as a function of the determinate model parameters by  introducing  additional dimensionless 
parameter: ${\bar A_0}= A_0 L_0/U_0$, and for brevity we drop the bar hereafter. Fig. 8a depicts 
the plot for the SNR as a function of $T$ for the  parameter values of $A_0=0.1$, $A=0$ and $U_0=2.0$. The parameter $N$ is varied from $N=1$ to $N=10$
 for a  variable $\gamma$ case.   The SNR exhibits 
monotonous noise strength dependence revealing a peak at an optimal noise strength $T_{opt}$. 
Our analysis  shows that $T_{opt}=U_{0}/3$ showing that  $T_{opt}$ increases as $U_0$ 
decreases. In Fig. 8b, the SNR as a function of $T$ is plotted  for the  parameter values of
$U_0=2.0$ and $L_{0}=1.0$ 
for a  constant $\gamma$ case and $A_0=0.1$. 
As shown in the figures, once again the  SNR exhibits 
 a peak at an optimal noise strength $T_{opt}$. Furthermore, as $N$ increases, $T_{opt}$ decreases and 
the SNR increases with $N$. This  numerical result clearly indicates that the abnormal calcium firing rate increases in the presence of  weak external signals since the first passage time considerably decreases in the presence of signals. Thus our study is vital in developing antiarrhythmic strategy.

\begin{figure}[]
\centering
\subfigure[ ] 
{
  \includegraphics[width=5cm]{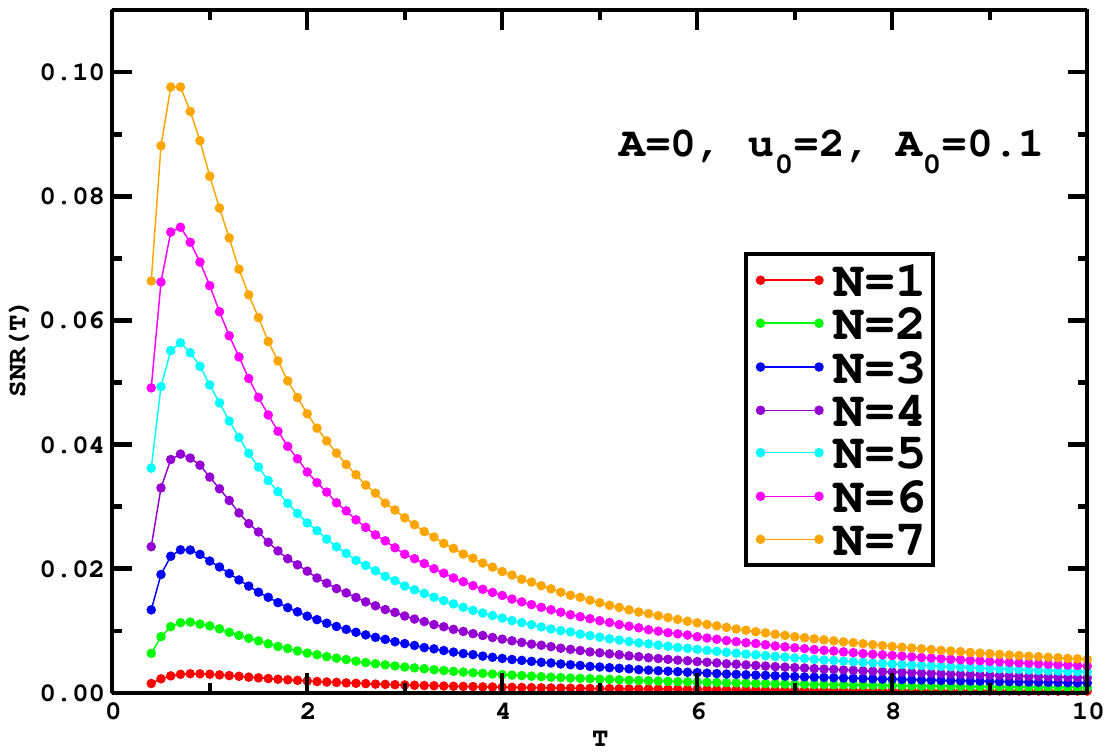}
}
\subfigure[ ] 
{
  \includegraphics[width=5cm]{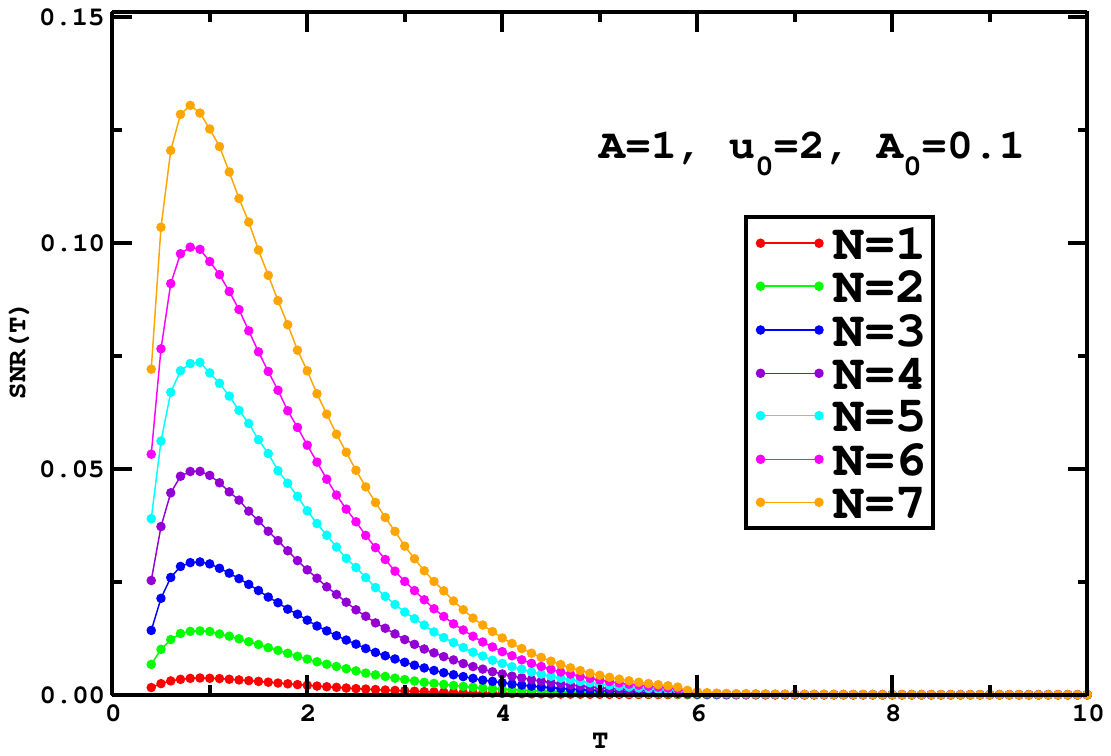}
}
\caption{ (Color online) The  $SNR$  as a function of $T$ for the  parameter values of
$A_0=0.1$, and $U_0=2.0$. Figures 8(a) and 8(b) are plotted for a constant $\gamma$  ($A=0$) and variable $\gamma$ ($A=1$) cases,
respectively.)
respectively.} 
\label{fig:SNR} 
\end{figure}

\section {Summary and conclusion}

In this work first we study the first passage time  of a single particle
both for temperature dependent and independent viscous friction cases. The 
the simulation results depict that the first passage time  is considerably smaller 
when $\gamma$ is temperature dependent.  In both cases the escape rate increases as 
the noise strength increases and  decreases as the potential barrier increases.  We 
then  extend our study for $N$ particle systems. The first passage time $T_{N}$ for 
one particle out of $N$ particles to cross the potential barrier can be studied 
via numerical simulation for any 
cases. It is found that $T_{N}$ is considerably  smaller when the viscous friction is 
temperature dependent. For both cases, $T_{N}$ decreases as the noise strength increases 
and as the potential barrier steps down.  In high barrier limit, $T_{N}=T_{s}/N$ 
where $T_{s}$ is the first passage time  for a single particle. In general as the number of particles
$N$ increases, $T_{N}$ decreases.
.

We also study our model in the presence of time varying periodic signal. In this case 
the interplay between noise and sinusoidal driving force in the bistable system may 
lead the system into stochastic resonance. Via numerical simulations we study how the signal to
noise ratio (SNR) behaves as a function of the model parameters. The SNR depicts a pronounced peak
at particular noise strength $T_{opt}$. The SNR is higher when $\gamma$ is temperature
dependent. For many particles system, SNR is considerably amplified as the number of particles $N$ 
steps up showing the weak periodic signal plays a vital role in controlling the noise induced
dynamics of excitable systems.

In conclusion, in this work we presented a very important model which helps to understand 
the dynamics of excitable systems such as neural and cardiovascular systems. Thus, this numerical
study is crucial not only for the fundamental understanding of excitable systems, but
also in providing a basic paradigm in to understand thermally activated transport features 
in various biological systems.

{\it Acknowledgment.\textemdash} 
We would like to thank Mulugeta Bekele for the interesting discussions we had. MA 
would like to thank Mulu Zebene for the constant  encouragement. 

\end{document}